\documentstyle[amssym,epsfig]{mn}

\newcommand{\aap}[2]{A\&A, #1, #2}

\newcommand{\apj}[2]{ApJ, #1, #2}

\newcommand{\mn}[2]{MNRAS, #1, #2}

\newcommand{\nature}[2]{Nature, #1, #2}
\title{The Importance of Fourier Phases for the Morphology of Gravitational
Clustering}

\author[Chiang] {Lung-Yih Chiang\thanks{E-mail:chiang@tac.dk}\thanks{http://www.tac.dk/{\em\~~}chiang} \\  Theoretical Astrophysics
Center, Juliane Maries Vej 30, DK-2100, Copenhagen,
Denmark.}

\date{Accepted 2000 ???? ???; Received 2000 ???? ???; in original form 2000 ???? ??}

\begin{document}

\maketitle

\begin{abstract}
The phases of the Fourier modes appearing in a plane-wave
expansion of cosmological density fields play a vital role in
determining the morphology of gravitationally-developed
clustering. We demonstrate this qualitatively and  quantitatively
using simulations. In particular, we use cross-correlation and
rank-correlation techniques  to quantify the agreement between a
simulated distribution and phase-only reconstructions. The
phase-only reconstructions exhibit a high degree of correlation
with the original distributions, showing how meaningful spatial
reconstruction of cosmological density fields depends  more on
phase accuracy than on amplitudes.

\end{abstract}


\section{Introduction}
The standard theory of the origin of the large-scale structure of
the universe involves the assumption that the structure we see
today grew by gravitational instability from initial small
fluctuations in the density field. In most popular variants of
this model, particularly those involving cosmic inflation, the
initial fluctuations are of a particularly simple form known as a
Gaussian random field. Gaussian random fields are useful because
many properties of Gaussian random density fields can be
calculated analytically (e.g. Bardeen et al. 1986). Some direct
motivation for such an assumption emerges from inflationary
models, wherein the fluctuations are generated by quantum
oscillations of the scalar field driving inflation (the
``inflation''). Even if inflation turns out to be incorrect,
however, the Central Limit Theorem tends to produce Gaussian
fluctuations in any linear process, so that they are the most
generic form for small initial conditions and a natural default
assumption.

One particularly interesting property of Gaussian random fields is
that the requirement for the density contrast
$\delta({\bf x})=[\rho({\bf x})-\rho_{0}]/\rho_{0}$ to be a Gaussian
random field is equivalent to that the real and imaginary parts of its
Fourier components $\tilde{\delta}_k$, where
\begin{equation}
\delta ({\bf x}) =\sum \tilde{\delta}({\bf k}) \exp(i{\bf k}\cdot
{\bf x}),
\end{equation}
are independently distributed. In other words, the Fourier modes
$\tilde{\delta}({\bf k})$,
\begin{equation}
\tilde{\delta}({\bf k})=|\tilde{\delta} ({\bf
k})|\exp(i\phi_{\bf_k}), \label{eq:fourierex}
\end{equation}
possess phases $\phi_{\bf k}$ which are independently distributed
and uniformly random on the interval $[0,2\pi]$. As the density
field is simply a sum over a large number of Fourier modes and if
the phases of each of Fourier modes are random, the Central Limit
Theorem guarantees  that the resulting superposition of the
one-point probability distribution ${\cal P}(\delta)$ is close to
Gaussian and that all of the field's joint probability
distributions are multivariate Gaussians. The statistical
properties of an isotropic Gaussian random field are then
completely specified by its second-order statistical quantity: the
covariance function, or alternatively, its power spectrum
$P(k)=\langle\tilde{\delta}^2(k)\rangle$.

In the framework of gravitational instability, the growth of
fluctuations can be understood analytically when the density
fluctuation amplitude is small compared to the mean density; the
linear perturbation theory tells us that each Fourier mode grows with
the same rate independent of wavenumber and the statistical distribution of
$\delta$ remains constant except its variance.

The linear theory breaks down when $\langle\delta^2\rangle$ is
comparable with unity or beyond, and different Fourier modes start
coupling.  One way to look at the mode coupling is that $\delta$
is always constrained to value $\delta \geq -1$. When the
perturbation is small, the tail of  ${\cal P}(\delta)$ in the part
of $\delta <-1$ assigned by Gaussian distribution is negligible,
because probability of $\delta<-1$ is small. When  the density
field evolves beyond the linear regime, i.e., $\sigma^2 \equiv
\langle \delta^2 \rangle \sim 1$, a long tail at high $\delta$ is
generated while lower bound is confined at $\delta=-1$. Gaussian
condition is therefore invalid, in that mode coupling effect
causes the initial condition to skew. Terms in the evolution








\begin{figure}
\caption{Visual demonstration of the importance of phase information
for clustering morphology. Plate (a), (c) and (e) have the same phase
configuration, so do plate (b) and (d). Plate (a), (d) and (f),
however, share the same power spectrum, or alternatively, two-point
correlation function, so do (b) and (c). See the text for
details. }\label{demo}
\end{figure}
\newpage

\noindent equations for the Fourier modes that represent coupling between
different modes are of second (or higher) order in $\delta$ and
these are neglected when first-order perturbation theory is
considered. Phases of Fourier modes therefore are therefore
coupled together in a way which is yet to be fully elucidated but
which has been recently investigated by Chiang \& Coles (2000) and
Coles \& Chiang (2000).

One of the reasons for studying Fourier phases in depth is the
question of possible primordial non-Gaussianity of the initial
density distribution of the Universe. For example, there have been
claims of non-Gaussianity from analysis results in the COBE DMR
sky maps of cosmic microwave background using a number of
different  diagnostics: bispectrum analysis (Ferreira et al.
1998); the wavelet transform (Hobson et al. 1999); and
Minkowski functionals (Schmalzing \& Gorski 1998).
The most direct practical approach, however, is through the
distribution of Fourier phases. This diagnostic can also avoid the
subtlety of more standard methods that depend on the distribution
function ${\cal P}(\delta)$(Scherrer et al. 1991), in that a
density field with Gaussian single-point density distribution
${\cal P}(\delta)$ is not necessarily a Gaussian field. Phase
information is also important as a statistical diagnostic of
non-linearity when the evolution of clustering in the non-linear
regime which is generally intractable analytically. Indeed, , as
we shall show, it is also closely related to the morphology of
gravitational clustering and through this to the dynamical origin
of structure.

The layout of this paper is as follows. In section 2 we present a
visual demonstration of the link between Fourier phases and
clustering morphology. Section 3 contains a brief discussion of
some of the properties of phases and some pitfalls that must be
avoided when using them as part of a clustering descriptor. In
Section 4 we introduce a cross-correlation parameter $S$ and a
rank-correlation $\tau$ as quantitative measures of the agreement
between two distributions. In Section 5 we display the results of the
correlation tests between sample distributions
and phase-based reconstructions, and between simulations evolving from
different initial power spectra but the same initial phase set. A
brief discussion of the results follows in Section 6.

\section{Visual Demonstration}
To give a qualitative, visual description of the key ideas in this
paper consider Fig.~\ref{demo}, in which we isolate the role of
phases in determining clustering morphology. Plates (a) and (b)
are two example realisations from two 2D N-body experiments, evolving
from different initial power-law power spectra and different initial
phase sets. Plate (a) is evolved
from power spectral index $n=-1$, and (b) from $n=1$. We perform a
Fourier transform on both realisations, e.g., $\delta ^{a}({\bf
x}) = \sum\tilde{\delta}^{a}_{\bf k} \exp(i{\bf k}\cdot {\bf x})$,
where $\tilde{\delta}^{a}_{\bf k}=|\tilde{\delta}^{a}_{\bf
k}|\exp(i\phi^{a}_{\bf k})$. Plate (c) is obtained by taking the
inverse Fourier transform from the combination of the phases of
Fourier modes from (a), and the amplitudes from (b), i.e., \({\cal
F}^{-1}[|\tilde{\delta}^{b}_{\bf k}|\exp(i\phi^{a}_{\bf k})]\).
Plate (d) is from the phases from (b) but amplitudes from (a).
Therefore, plate (a) and (c) share the same phase configuration,
so do (b) and (d). It is easy to see resemblance between (a) and
(c), and between (b) and (d). Note that plate (a) and (d) have the
same power spectrum, or equivalently, two-point correlation
function, as do plate (b) and (c). Plate (e) is the inverse
Fourier transform from only the phases from (a): \({\cal
F}^{-1}[\exp(i\phi^{a}_{\bf k})]\); it therefore retains only the
phase information from (a). In plate (f), each mode keeps the same
amplitude so its power spectrum is unchanged (i.e. the same as
plate (a) and (d)) but the phases are redistributed randomly among
the modes before inverse Fourier transform. Again,  plate (e), the
phase-only reconstruction (hereafter phase-only reconstruction),
resembles the original distribution plate (a), and plate (c), the
same-phase amplitude-swapped reconstruction (hereafter
amplitude-swapped reconstruction), but (f) which has the same
power spectrum as (a) but random phases (hereafter random-phase
reconstruction), is featureless. This experiment suffices to show
very clearly how phases determine morphology.

Another interesting property of Fourier phases in clustering
morphology is that two realisations evolving from the same initial
random phase set, though different power-law power spectra, will have
their extrema at the same locations. The fact that Plate (a) and (b)
are evolved from different initial phase sets is for demonstration purpose.

\section{Quantifying Phase Shifts}
Fourier phases reflect the locations of spatial `events'
more than Fourier amplitudes do. For a single hypothetical spike,
represented by the Dirac $\delta$-function $\delta_D(x-x_0)$, the
amplitudes are constant and phases are $k x_0$. This has a
white-noise spectrum, but very strong phase correlation. The phase
configuration of a  spike is very similar to that  of a single
density peak evolved from 1D Zel'dovich approximation; see Chiang
\& Coles (2000) for details.  A translation $x^{'}$ of the Dirac
$\delta$-function density field, $\delta_D(x-x_0-x')$, has no
effect on the Fourier amplitudes, but the phases now become
$k(x_0+x')$, which  suffers a shift by a linear term proportional
to wave number $k$. This dependency of phases on the choice of
origin means that some care must be taken when trying to extract
meaningful information.  Some  previous studies focused on the
evolution of individual phases away from their initial values
(Ryden \& Gramann 1991; Soda \& Suto 1992; Jain \& Bertschinger
1998). The mean deviation from the initial phase can be defined as
\begin{equation}
\Delta\phi(k,a)=\langle | \Delta \phi({\bf k},a) | \rangle  =\langle |
\phi({\bf k},a)-\phi({\bf k},a_i) | \rangle,
\end{equation}
where the averages are performed over the different modes within a
shell in k-space  whose wave numbers lie in the range $k-0.5 < |{\bf
k}| < k+0.5$. As long as there are enough modes, the maximal value of
$\Delta\phi(k,a)$ is $2\pi/3$ (Jain \& Bertschinger 1998), which can
be understood as a variable obtained from the change from a random field
to a hypothetical spike, i.e., from random phases to a very
ordered state, $\phi(k,a)=k x_0$( mod $2\pi$), where $x_0$ is the
location of the spike and takes on any values except
zero. This statistic will change following the translation of
`events', e.g., $\Delta\phi(k,a)=\pi$ if the hypothetical spike is
shifted to the origin.

\section{Cross and Rank Correlations}
Although the examples displayed in Section 2 serve to demonstrate
the importance of phases, it is necessary to measure objectively
how well the phase-based reconstructions compare not only with
the original distribution, but also with reconstruction from, say,
random phases. The tool we use here is the cross-correlation
statistic introduced by Coles, Melott \& Shandarin (1993). A
correlation coefficient is defined as
\begin{equation}
S_{\delta
\delta^{r}}=\frac{\langle(\delta_{ij}-\overline\delta)(\delta^{r}_{ij}-\overline\delta^{r})\rangle}{\langle(\delta_{ij}-\overline\delta)^{2}\rangle^{1/2}\langle(\delta^{r}_{ij}-\overline\delta^{r})^{2}\rangle^{1/2}}\;\;,
\label{eq:xcorr}
\end{equation}
where $\delta_{ij}$ and $\delta^{r}_{ij}$ represent two density
distribution and $\overline\delta$ and $\overline\delta^{r}$ are
their mean densities, respectively. The indices $i$ and $j$ label
the pixel positions in the two-dimensional simulations we use here
for illustration. The parameter $S$ compares the density value of
each grid point at $(i,j)$ in the original distribution $\delta$
with the corresponding grid point of the reconstruction
$\delta^{r}$; averages are taken over all grid points.
$S_{\delta\delta^{r}}=1$ denotes `completely positive
correlation', a perfect agreement between $\delta$ and
$\delta^{r}$, and a value of zero indicates the two distributions
are uncorrelated.  What is useful about this test is that it
compares the morphology between two distributions, point by point, but
does not take into account their variances.  This is because
$S_{\delta\delta^{r}}=1$ when $\delta=\,C \delta^{r}$, $C$ being a
constant. If the structures of two distributions are different,
$|S|$ is less than unity.

Cross-correlation tests the spatial correspondence of the
locations of `events', such as clusters of points or edges,
between two distributions. Even the comparison by a small
displacement between two identical periodic distributions will
result in low value of $S$. It should be pointed out, therefore,
that this grid-by-grid test is severe. As we have explained,
phases are related to preservation of locations of `events'.
Relative magnitudes of the `events', however, are not preserved when
the information of Fourier amplitudes is partially or totally
lost. To relax the test, comparisons are made not only between raw
density distributions $\delta$ and $\delta^{r}$, but also between
smoothed distributions $\delta({\bf x},R)$ and $\delta^{r}({\bf
x},R)$. For this purpose, a Gaussian window function is chosen to
smooth the field,
\begin{equation}
\delta({\bf x},R)=\int d^{2}{\bf x^{'}} \: \delta({\bf x^{'}}) \:
(\sqrt{2\pi}R)^{-2} \exp(-\frac{|{\bf x}-{\bf x^{'}}|^{2}}{2R^{2}}). \label{eq:window}
\end{equation}

As well as the linear cross-correlation $S$, we also use a
non-parametric rank-correlation, Kendall's $\tau$
parameter(Kendall and Gibbons 1990). This parameter uses the
relative ordering of ranks to measure the degree of agreement
between two compared distributions. For any two grid points
$(i,j)$ and $(i^{'},j^{'})$, the given value is $+1$ if the
relative ordering of the density values of $\delta(i,j)$ and
$\delta(i^{'},j^{'})$ is the same as  $\delta^{r}(i,j)$ and
$\delta^{r}(i^{'},j^{'})$, i.e., either both decreasing or both
increasing from $(i,j)$ to $(i^{'},j^{'})$; $-1$ when one pair is
increasing and the other is decreasing. The $\tau$ parameter
compares $N(N-1)/2$ pairs from total $N^2$ grid points and is
normalized to $[-1,1]$. Rank correlation measures such as this do
not look specifically for linear association between the image and
reconstruction, but for one-to-one ordering of the values in one
relative to those in the other. If there is strong non-linear
association, then $\tau$ can be close to unity even though $S$ may
be small. Kendall's $\tau$ parameter is used here as an auxiliary
test, which is again ``softened'' with a Gaussian window function
as in eq.~(\ref{eq:window}).

\begin{figure}
\begin{center}
\epsfig{file=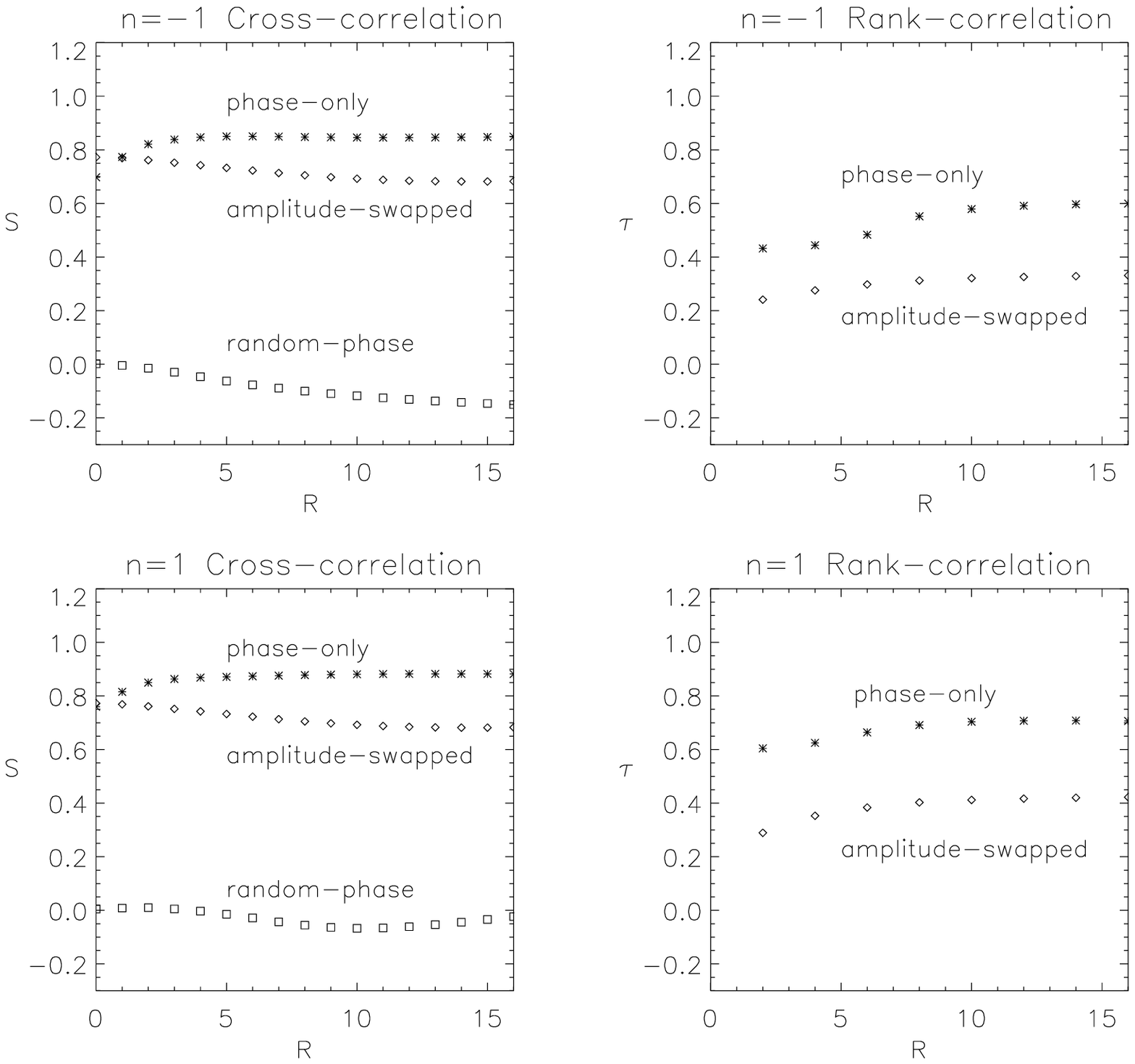, width=8cm} \caption{Plots of
cross-correlation coefficient $S$ and rank-correlation coefficient $\tau$ against smoothing scale $R$ (in
computer grid units) for the comparison between sample distributions
and its phase-based reconstructions. Phase-only and amplitude-swapped
reconstructions retain the morphology of the original structure of the
sample realisation, which is shown by a high degree of
cross-correlation. Random-phase reconstruction, however, bears no
resemblance to the original structure, hence low $S$. Rank-correlation
compares the relative magnitudes of the density peaks. Due to the loss of the
information on the Fourier amplitudes in the reconstructions, the
relative ordering of the ranks of the density magnitudes is partially
disturbed. The phase-only reconstructions keep the ordering better
than amplitude-swapped reconstructions. }\label{corr}
\end{center}
\end{figure}

\section{Results}
Fig.~\ref{corr} shows cross-correlation coefficient $S$ and
rank-correlation coefficient $\tau$ drawn against smoothing scale
$R$ in computer grid units. Each panel includes correlations
between the sample distribution and phase-only, amplitude-swapped
reconstructions. Correlations of random-phase reconstructions are
calculated only in $S$. On the top-left panel of Fig.~\ref{corr},
the sample distribution evolved from spectral index $n=-1$(Fig.1a)
is compared with phase-only reconstruction(Fig.1e), and with
amplitude-swapped reconstruction (Fig.1c). The random-phase
reconstruction (Fig.1f) is also compared for reference. The sample
distribution of lower-left panel is evolved from $n=1$(Fig.1b).
Even before smoothing, the higher degree of cross-correlation from
phase-only and amplitude-swapped reconstructions than from
random-phase reconstruction shows the ability of phases to retain
the morphology of the sample distributions. Non-resemblance of
random-phase reconstructions is shown through low value of $|S|$.
That smoothing of the reconstructions increases (for
amplitude-swapped it even decreases) certain level of $S$
indicates the limitation of phases on reconstruction with the loss
of amplitude information.

\begin{figure}
\begin{center}
\epsfig{file=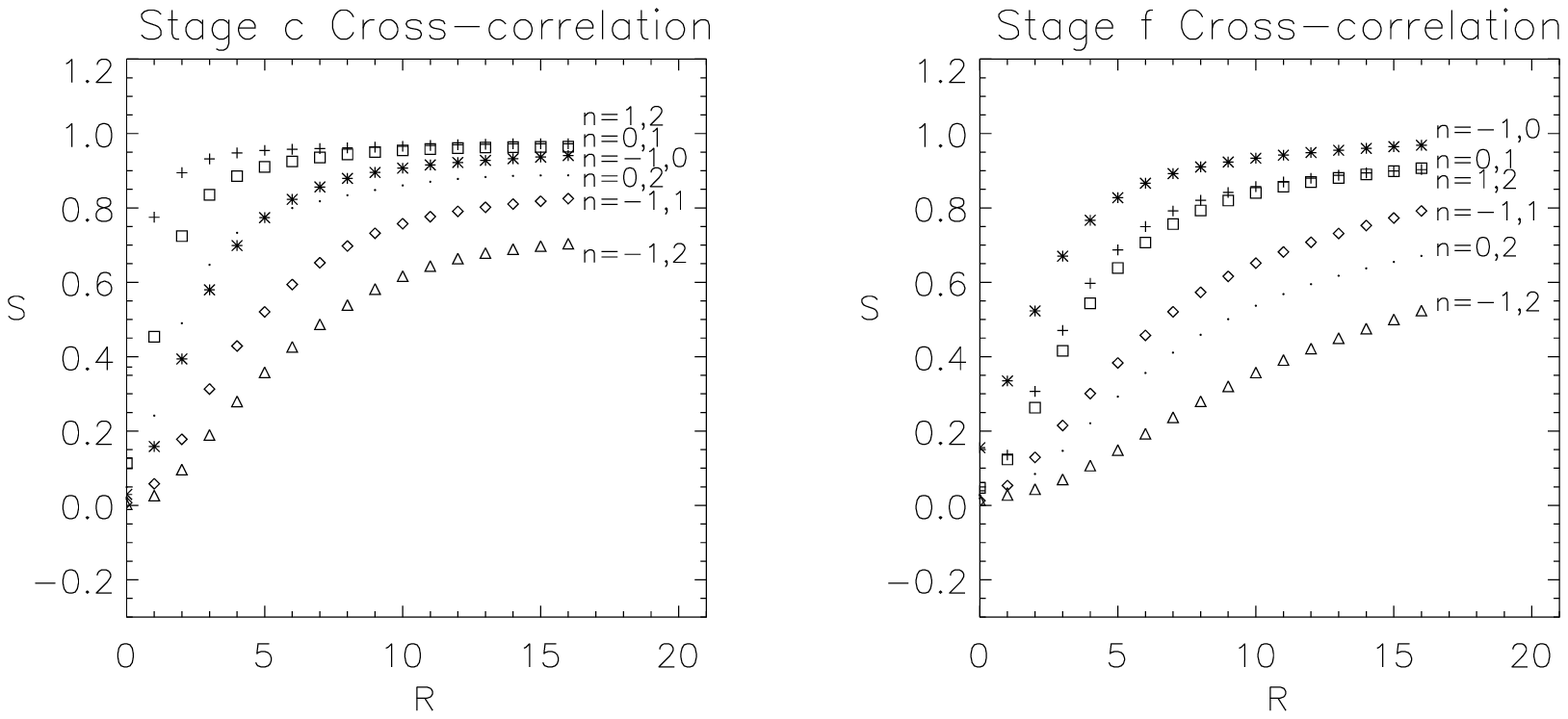, width=8cm} \caption{Comparison of morphology
of two different stages in N-body simulations in terms of
cross-correlation coefficient $S$ between realisations evolving from
different initial power-law power spectra but the same initial phase
set. The plots are drawn against smoothing scale $R$, as in
Fig.~(\ref{corr}).}\label{xcorr1}
\end{center}
\end{figure}

The rank-correlations, on the other hand, show that phase-only
reconstructions have higher agreement than amplitude-swapped ones on
relative density magnitudes, which is due to the `twist' on the
latter's power spectra. Surely, the loss of the Fourier amplitude
information destroys part of the signal, as shown in the
rank-correlation of the right panels. On the morphology of
gravitational clustering, however, phases play a much more important
role than amplitudes.

It emphasizes this point still further to test the correspondence
between simulations evolving from different initial power spectra
but the same initial phases. In Fig.~(\ref{xcorr1}) we compare the
morphology for some relevant examples in terms of
cross-correlation coefficient $S$ deployed above. In particular,
we show  realisations obtained by evolving $N$-body experiments
from different 2D initial power-law power spectra $n=-1$, $0$ $1$
and $2$. The left panel is the comparison of stage $c$ in which
the scale of non-linearity is $k_{NL}=64 k_{f}$, and the right
panel is that of later stage $f$ with $k_{NL}=8 k_{f}$(Chiang \&
Coles 2000, Beacom et al. 1991). We use the fundamental mode
$k_{f}=2 \pi/L$ as length unit, where $L$ is the length of the
side of the simulation square.

Before smoothing, there is little or no correspondence between any
of the realisations. This is due to the severe point-by-point test
of the cross-correlation. After smoothing, however, there is a
dramatic improvement between in the correspondence between $n=-1,
0$, $n=0, 1$, and between $n=1, 2$. We can examine the morphology
evolving from the same phase set by the characteristic scale
$k_{NL}$. For example, $k_{NL}$ for stage $c$ corresponds to 8
computer grid units, thus the smoothing on scales beyond 8 grid
units erases non-linearities, while the  larger scale structure
remains in the linear regime. The linear growth of the density
fluctuations set up by the same initial phases depends only on
time. The comparison therefore indicates the intrinsic difference
in clustering morphology arising from different initial power
spectra. What we see on the scales beyond $k_{NL}$ is that the
correspondence is low between $n=-1$ and $n=1, 2$.

The correspondence deteriorates when the difference between the
spectral indices increases, i.e., the difference in intrinsic
morphology is significant, or when the evolution goes into highly
non-linear regime, where particles move away from their initial
Lagrangian position and interact with each other non-linearly.
Simulations evolving from the same initial phase configuration
tend to develop nonlinear structures at or near the same spatial
locations, but these structures appear with different contrast
when the initial spectra are different. For example, filaments
appear in the nonlinear regime in all cases, but for spectra with
large $n$ these tend to be less well defined and broken up into
clumps. Our statistic $S$ takes into account both the position and
amplitude of structures that form so it indicates a deteriorating
agreement for very different spectra. Nevertheless, it is clear
that there is strong imprint upon the morphology of the initial
phases resulting from the process of gravitational clustering.
This can also be seen visually in the pictures shown by Beacom et
al. (1991).

\begin{figure}
\begin{center}
\epsfig{file=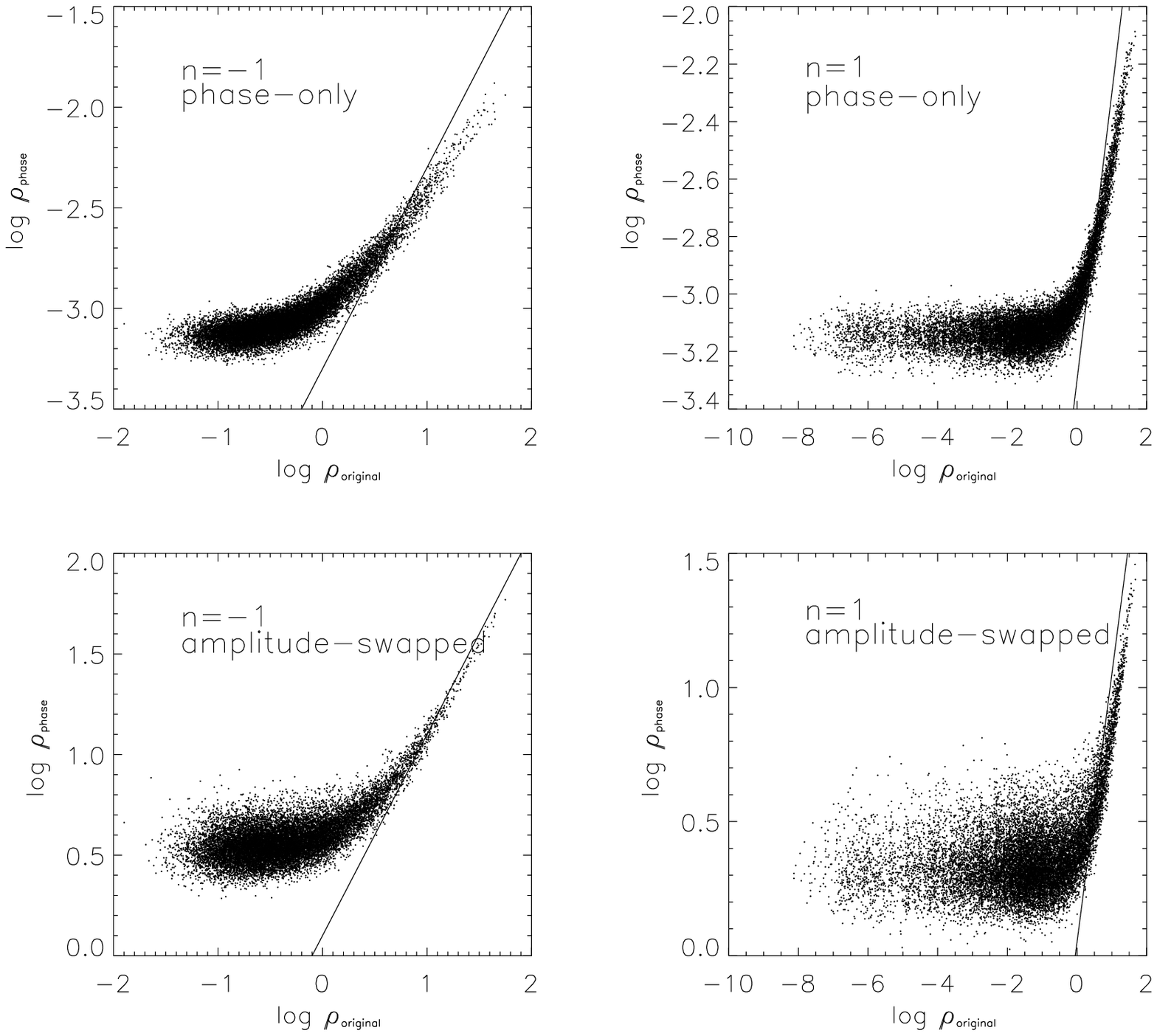,width=8cm}
\caption{Scatter diagrams of cell
density between smoothed original distributions and
reconstructions. The smoothing scale is two grid units to produce a
continuous distribution. In each panel
 only one in four grid points in each spatial direction  on the
$512^2$  grid are sampled: 16384 points and a straight line with
slope equal to unity is added for reference. The top-left panel is a
comparison between Fig.1(a) and Fig.1(e) , the bottom-left is between
Fig.1(a) and (c), and the bottom-right is between (b) and (d).}
\label{scatter}
\end{center}
\end{figure}

The cross-correlation coefficient being a single number, so one
cannot infer from it precisely how the phase-based reconstructions
perform relative to the morphology of the original distribution.
One intriguing question is how the density regions from
reconstructions can be compared to those from the original
distributions. Is it $\delta \propto \delta^{r}$ that produces
high value of $S$?  From rank-correlation coefficient $\tau$, this
is is not the case. In order to get more information between the
two distributions, we also made a grid-by-grid comparison between
them. A scatter diagram can be drawn in logarithmic scales for the
comparison between each cell density from phase-based
reconstructions against the corresponding one from the original
distribution. If  $\delta \propto \delta^{r}$, the points  scatter
along a straight line of slope equal to unity, and the relative
magnitude of `events' is preserved by a linear scale factor.
Fig.~\ref{scatter} shows the scatter diagram between original
distribution and phase-only, amplitude-swapped reconstructions for
$n=1$ and $n=-1$. Not all the points scatter along the straight
line, as is suggested. Instead, most points scatter horizontally,
particularly for phase-only cases. Also notice also that, in both
reconstruction cases, the density values span only roughly one
order of magnitude and the points that do not align well
correspond to very small fluctuations. High values of $S$ result
from phase configuration preserving the locations of high-density
`events' with large magnitudes in the original structure. The
relatively flat phase-only reconstructed density distribution is
caused by the flat power spectrum, in which the Fourier amplitude
for each mode is squashed into unity. The lower two panels are
scatter diagrams of the amplitude-swapped reconstructions against
the originals, i.e., in Fig.~\ref{demo} between (c) and (a) and
between (d) and (b). The difference is amplitude-swapped
reconstructions have the power spectra from the alternative sample
distributions. The $n=-1$ amplitude-swapped reconstruction has the
power spectrum from the original distribution of $n=1$, which
gives more power on small scales in the reconstruction, the $n=1$
reconstruction, on the other hand, has more power on large scales.
This adverse effect causes the points of the low fluctuations to
spread on the scatter diagram, which, however, doesn't decrease
much the cross-correlation coefficient $S$ between phase-only and
amplitude-swapped reconstructions. This is in accord with with our
visual impression that our eyes pick up the maxima between the
distributions for comparison.

In Fig.~\ref{scatter1} the scatter diagrams for the realisations
evolving from the same phase set are also produced. Here only two cases
are chosen, i.e., between $n=-1$ and $0$, and between $n=-1$ and
$2$. The left two panels are comparison between $n=-1$ and $n=0$, and
the scattering is expected, which nonetheless follows the straight
line.  For the right panels, at early stage $c$, the deviation from
the straight line is more isotropic, hence the correspondence is low.  At late stage $f$, there is even
more deviation, which comes from the mapping from the low-density regions
of stage $f$, $n=2$. The reason is as follows. For the realisation of
$n=2$ to reach the same level of non-linearity, the variance is higher
than that of $n=-1$, and there are a substantial number of
voids. After smoothing to create continuous distribution, the
void regions are smoothed as low-density regions, which is seen in the
bottom-right panel.

\begin{figure}
\begin{center}
\caption{Scatter diagrams of cell
density between simulations evolving from the same phase set. Here we
choose between spectral index $n=-1$ and $n=0$, and between  $n=-1$ and
$n=2$. The smoothing scale is four grid units. As in
Fig.~\ref{scatter}, 16384 points are sampled and a straight line with slope equal to unity is added for reference. }
\label{scatter1}
\end{center}
\end{figure}

\section{Conclusion}
We have shown the importance of Fourier phases on morphology
 of gravitational clustering via qualitative and quantitative demonstrations.
 It is interesting to remark upon the similarity of the results we
 have obtained here and those presented in Coles et al. (1993).
 The latter authors were interested in the ability of simple
 analytic methods to reproduce the clustering displayed by full
 N-body computations. They showed in particular that the
 Zel'dovich approximation (Zel'dovich 1970) could
 reproduce the full numerical results quite well, with
 cross-correlations similar to those we have found here. The
 Zel'dovich approximation works as well as it does in this respect
 because it places the caustic surfaces forming sheets and
 filaments near to the correct location in the N-body experiment.
 In other words, the Zel'dovich approximation has a high phase
 fidelity. It is less good at getting the amplitudes right.
 As we have shown, however, phases dominate the morphology.

 With a new colour representation technique
(Coles \& Chiang 2000), phase information can be used to
distinguish between non-Gaussianity induced by gravitational
clustering and that by other mechanisms. For example, in the web
page
\begin{verbatim}
http://www.nottingham.ac.uk/~ppzpc/phases/cmb.html
\end{verbatim}
the phase configuration of the non-Gaussian temperature
fluctuations on the CMB sky induced by cosmic strings can be seen
to be intrinsically different from that of hierarchical clustering
developed in N-body simulations. A new algorithm based on the
analysis of the phase distribution of Fourier components is
recently devised to extract noise from point sources from the CMB
signal(Naselsky et al. 2000), which is another example practice
of the close link between Fourier phases and morphology.

Future large-scale galaxy redshift surveys and microwave sky maps
will reveal much morphological information about large-scale
structure in the Universe. Existing statistical technology,
however, is still dominated by second-order methods that are blind
to phase information. Our success in extracting this information
will therefore depend on the development of statistical methods
sufficiently sensitive to the key ingredient: the distribution of
Fourier phases.

\section*{Acknowledgments}
I thank Peter Coles for useful suggestions. This work
was partly supported by Danmarks Grundforskningsfond through its
support for TAC.


\begin{thebibliography}{}
\bibitem[]{bbks}   Bardeen J.M., Bond J.R., Kaiser N., Szalay A.S.,
                     1986, \apj{304}{15}
\bibitem[]{beacom} Beacom J.F., Dominik K.G., Melott A.L., Perkins S.P.,
                     Shandarin S.F., 1991, \apj{372}{351}
\bibitem[]{chiang}   Chiang L.-Y., Coles P., 2000, \mn{311}{809}
\bibitem[]{coles1}   Coles P., Chiang L.-Y., 2000, \nature{406}{376}
\bibitem[]{coles2}   Coles P., Melott A.L., Shandarin S.F., 1993,
                     \mn{260}{765}
\bibitem[]{fmg}    Ferreira P.G., Magueijo J., Gorski K.M.,
                   1998, \apj{503}{L1}
\bibitem[]{hobson} Hobson M.P., Jones A.W., Lasenby A.N., 1999 \mn{309}{125}
\bibitem[]{jain}   Jain B., Bertschinger E., 1996, \apj{456}{43}
\bibitem[]{jain2}  Jain B., Bertschinger E., 1998, \apj{509}{517}
\bibitem[]{kendall}Kendall M., Gibbons J.D., 1990, Rank Correlation
                      Methods, Oxford University Press, New York
\bibitem[]{naselsky}Naselsky P., Novikov D., Silk J., astro-ph/0007133
\bibitem[]{ryden}  Ryden B.S., Gramann M., 1991, \apj{383}{L33}
\bibitem[]{scherrer}Scherrer R.J., Melott A.L., Shandarin S.F.,
                      1991, \apj{377}{29}
\bibitem[]{schmalzing}Schmalzing J., Gorski K.M., 1998, \mn{297}{355}
\bibitem[]{soda}   Soda J., Suto Y., 1992, \apj{396}{379}
\bibitem[]{zel}    Zeldovich Y.B., 1970, \aap{5}{84}

\end{thebibliography}
\end{document}